\documentclass[a4paper,twocolumn,english,10pt, prb, aps]{revtex4-1}
\usepackage{lmodern}
\usepackage[T1]{fontenc}
\usepackage[latin9]{inputenc}
\usepackage{amsmath}
\usepackage{amssymb}
\usepackage{graphicx}

\makeatletter

\pdfpageheight\paperheight
\pdfpagewidth\paperwidth

\makeatother

\usepackage{babel}
\begin{document}

\title{Optical remote control of a single charge qubit}

\author{T. S. Santana}
\email{sil.ted.s@gmail.com}

\affiliation{Departamento de F\'{i}sica, Universidade Federal de Sergipe, 49100-000, Brazil}

\author{F. A. G. Almeida}
\email{assis.almeida.se@gmail.com}

\affiliation{Departamento de F\'{i}sica, Universidade Federal de Sergipe, 49100-000, Brazil}

\date{\today}
\begin{abstract}
Both the electron transport-based qubits, implemented through double quantum dots, and the sources of indistinguishable single photons like self-assembled quantum dots are strong candidates for the implementation of quantum technologies, such as quantum computers and quantum repeaters. Here, we demonstrate a reliable way of coupling these two types of
qubits, uncovering the possibility of controlling and reading out the population of the double quantum dot via optical excitation. It is also shown that, in spite of the decoherence mechanisms affecting the qubits, the entanglement between them is achievable and, consequently, the implementation of the suggested system in quantum technologies is feasible.
\end{abstract}
\maketitle

\section{Introduction}

The entanglement between two quantum systems was noticed by Einstein, Podolsky, and Rosen when they discussed the validity of the quantum mechanics in its early times \cite{EPS_paradox_1935}. At that period, it was pointed out that the measurement in one of the systems could lead to instantaneous information about the state of the other quantum system, even at a distance \cite{schrodinger_1935,bell2004speakable},
although the information cannot travel faster than the speed of light. Actually, the entanglement between two or more qubits is a crucial ingredient for quantum computation, quantum information and quantum communication \cite{vedral2006_book,nielsen2002quantum}. However, in nature, the physical qubits always interact with some kind of reservoirs, which mitigates the quantum correlation. In this case, some distillation protocol has to be applied in order to recover the entangled state \cite{distillation_1,distillation_2,distillation_3,distillation_4,distillation_5,distillation_6,distillation_7}.

Semiconductor devices became strong candidates for the realization of quantum technologies, such as quantum computers \cite{loss1998quantum}, quantum memories \cite{memory_kroutvar2004} and quantum repeaters
\cite{repeater_wang2012}. The implementations are based on their optical properties, like in the case of self-assembled quantum dots (QDs) \cite{RF_vamivakas2009}, on the electron transport or on the electron spin, as in the case of double QDs defined by lithography techniques \cite{DQD_hayashi2003_1,DQD_ono2002_2,DQD_gorman2005_4}.

If the noise originated from the density charge fluctuation in the vicinity of the QDs is irrelevant for its dynamics, the coherence time of the charge qubit based on the electron position in a double QD can be in the order of 200 ns with a relaxation time to the ground state of about $100$ $\mu$s \cite{DQD_gorman2005_4} -- much greater than the typical lifetime of excitonic states in optically active QDs, which is typically between hundreds of picoseconds and a few nanoseconds \cite{coherence_englund2005,coherence_hennessy2007,coherence_kuhlmann2015,coherence_nguyen2011,coherence_2D_flatten2018}
-- allowing several qubit operations to be performed before the quantum features of the system is completely lost. Moreover, high coherence and indistinguishability of successive single photons have
been reported for different kinds of solid-state photonic devices \cite{HOM_claudon2010,HOM_gschrey2015,HOM_laurent2005,HOM_liu2018,HOM_patel2010,HOM_santori2002,HOM_Ted_2017},
which have already been utilized towards the implementation of non-universal quantum computers \cite{BS_spring2012,BS_wang2017}.

In this work, we suggest the construction of a bipartite system composed by a single-charge qubit and an optically active two-level system emitting single photons, which may be achieved from the application of lithography techniques in optically active photonic devices such as QDs in semiconductor chips and direct band-gap two-dimensional
transition dichalcogenide materials like WSe$_{2}$ and MoSe$_{2}$. The interaction between the charge qubit and the excited state of the optical QD via Coulomb energy enables the optical remote control of the first. Moreover, it is possible to obtain an entangled state between the emitter and the charge qubit, suggesting a new physical system for quantum communication purposes, which requires the conversion between stationary and flying qubits and the guiding of the photons emitted \cite{criteria_divincenzo2000}.

This paper is organized as follows. In Sec. II, we present the physical description of the system, while the results are exposed in Sec. III. In Sec. IV, we draw out our conclusions.

\section{Model}

\begin{figure}[th]
\begin{centering}
\includegraphics[width=\linewidth]{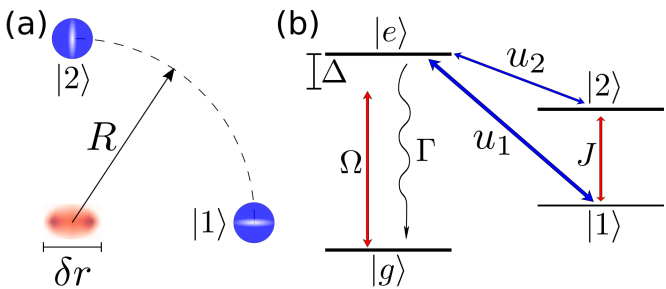}
\par\end{centering}
\caption{(a) Sketch of the system composed by an emitter surrounded by two electron traps away by the same distance $R$. (b) The optically active QD is driven from its ground state $|g\rangle$ to its excited state $|e\rangle$ through an excitation laser with frequency detuned by $\Delta$ from the optical transition. The coupling energy between the laser field and the exciton dipole moment is $\hbar\Omega$ and
the emitter suffers radiative decay with a rate equal to $\Gamma$. The single electron of the system can hop between the traps with a tunneling rate equal to $J$, changing the Coulomb energy $\hbar u_{m}$.
\label{fig:Fig1}}
\end{figure}

The system consists of two QDs distributed along a hemisphere of radius $R$ with a single-photon emitter in its center, as depicted in Fig. \ref{fig:Fig1} (a). The emitter QD is modeled as an optically driven two-level system in the dipole approximation \cite{mollow1969power}, where the Rabi frequency is $\Omega$ -- in this work only
continuous wave excitation is considered -- the detuning between the optical transition and the excitation field is $\Delta$ and the radiative decay rate is $\Gamma$. When the emitter is in the ground state, there is no interaction between itself and the single electron occupying one of the traps. On the other hand, if the central QD in driven to the excited state, the Coulomb interaction between
the trapped electron and the exciton may be significant, depending on the position of the electron trap relative to the radiating dipole. Allowing the electron to tunnel between the traps with a tunneling rate equal to $J$ and assuming that the ionization energies for all the electron traps are equal, the Hamiltonian of the system under the rotating wave approximation is
\begin{align}
\frac{H}{\hbar} & =\left(\Delta+u_{1}\right)\sigma^{\dagger}\sigma d_{1}^{\dagger}d_{1}+\left(\Delta+u_{2}\right)\sigma^{\dagger}\sigma d_{2}^{\dagger}d_{2}\nonumber \\
 & +\frac{\Omega(t)}{2}\sigma_{x}-J\left(d_{1}^{\dagger}d_{2}+d_{2}^{\dagger}d_{1}\right),\label{eq:H}
\end{align}
where $\sigma^{\dagger}$ ($\sigma$) is the creation (annihilation) operator for the emitter, acting on the Hilbert subspace composed by the state $|g\rangle$ and $|e\rangle$, $\sigma_{x}=\sigma^{\dagger}+\sigma$ is the Pauli matrix, and $d_{n}^{\dagger}$ ($d_{n}$) is the creation
(annihilation) operator for the $n$-th electron trap, acting on the Hilbert subspace with states $|1\rangle$ and $|2\rangle$, as depicted in Fig. \ref{fig:Fig1} (b). Considering that the average length of the radiating dipole $\delta r$ is much smaller than the distance $R$, the Coulomb energy is $\hbar u_{n}=U_{e}(R)\delta r\cos(\theta_{n})/R$,
where $U_{e}(r)$ is the Coulomb energy of two electrons separated by a distance $r$ and $\theta_{n}$ is the angle between the $n$-th trap position vector from the emitter and the vector $\delta\vec{r}$. Since we chose the second electron trap positioned at $\theta_{2}=\pi/2$ [Fig. \ref{fig:Fig1} (a)], only the first trap will have a nonzero
Coulomb energy.

The radiative decay of the emitter is described by the Lindblad superoperator, given by \cite{Lindblad_breuer2002}
\begin{equation}
L(\sigma)\rho=\frac{\Gamma}{2}\left(2\sigma\rho\sigma^{\dagger}-\sigma^{\dagger}\sigma\rho-\rho\sigma^{\dagger}\sigma\right),\label{eq:L}
\end{equation}
and the master equation of the system is
\begin{equation}
\frac{d\rho}{dt}=-\frac{i}{\hbar}\left[H,\rho\right]+L\left(\sigma\right)\rho,\label{eq:ME}
\end{equation}
where $\rho$ is the density matrix of the system.

For the initial state of the system, we consider that the emitter is in its ground state $|g\rangle$, while the probability of finding the electron in any of the traps is $50\%$. All the off-diagonal elements of the density matrix are zero at $t=0$. The trajectory of every element of the density matrix was obtained by numerically solving the system of differential equations [Eq. \ref{eq:ME}] using the fourth-order Runge-Kutta method \cite{numerical_press2007} with error tolerance of $10^{-6}$. The steady state density matrix $\rho^{s}$ is determined from the point where $t\gg1/\Gamma$ and $d\rho/dt\approx0$.

The charge noise is the main dephasing mechanism affecting the coherence of the electron position state in the charge qubit, but it is not considered here. We assume that its rate is much slower than the radiative decay and many qubit operations can be performed in the coherence time \cite{DQD_gorman2005_4}. The exciton-phonon coupling may decrease the efficiency of the remote preparation of the charge qubit, as well as the entanglement between the qubits, but, for the sake of simplicity, it was neglected here.

\section{Results}

\begin{figure}[th]
\begin{centering}
\includegraphics[width=\linewidth]{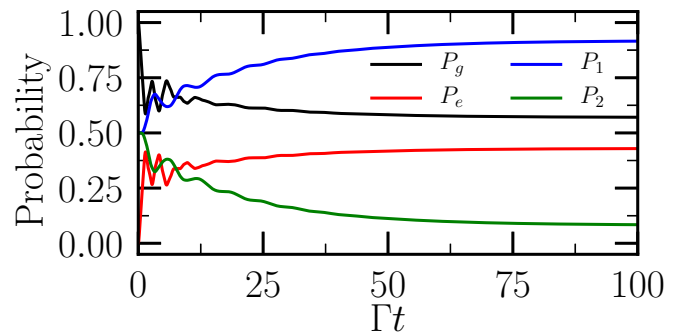}
\par\end{centering}
\caption{Temporal evolution of the states population of the system composed by an optically active two-level system with states $|g\rangle$ and $|e\rangle$, and a charge qubit based on the electron position states $|1\rangle$ and $|2\rangle$. The parameters used were $\Gamma=0.5$ GHz, $\Omega/\Omega_{sat}=3$, $J/\Omega_{sat}=0.5$, $u_{1}/\Omega_{sat}=9$, $u_{2}=0$ and $\Delta/u_{1}=-1$.\label{fig:Fig2}}
\end{figure}

It was observed that, when the parameters are adjusted to have $|u_{1}-u_{2}|>\Omega>J$ and the detuning $\Delta$ tuned to compensate the energy shift caused by the presence of the electron in one of the traps ($\Delta/u_{1}=-1$ or $\Delta/u_{2}=-1$), the evolution of the dipole states presents the typical Rabi oscillations, while the electron in the charge qubit is attracted to the trap which leads to resonance between the excitation field and the optically active two-level system, favoring the optical remote control of the charge qubit. This dynamics is depicted in Fig. \ref{fig:Fig2}, with $P_{i}=\langle i|\rho^{s}|i\rangle$ being the probability of finding the system in the state $|i\rangle$ when $\rho=\rho^{s}$, for $\Gamma=0.5$ GHz, $\Omega/\Omega_{sat}=3$, $J/\Omega_{sat}=1/2$, $u_{1}/\Omega_{sat}=9$, $u_{2}/\Omega_{sat}=0$ and $\Delta/u_{1}=-1$, where $\Omega_{sat}=\Gamma/\sqrt{2}$ is the saturation Rabi frequency
for a single two-level system \cite{rabi_knight1980}.

\begin{figure}[th]
\begin{centering}
\includegraphics[width=\linewidth]{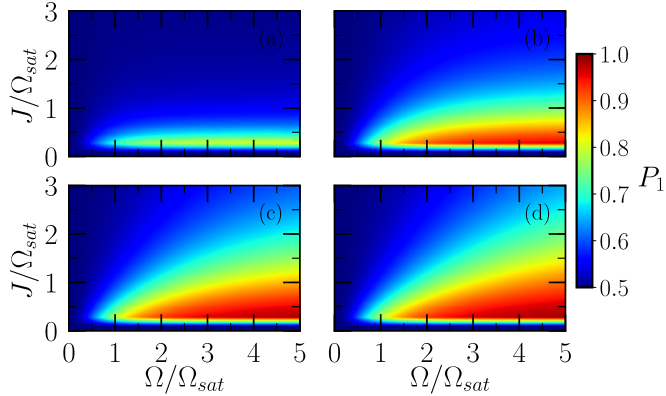}
\par\end{centering}
\caption{Probability of finding the electron in the first trap after the system reaches the steady state as a function of the Rabi frequency $\Omega$ and the tunneling rate $J$, both relative to the saturation Rabi frequency $\Omega_{sat}$, for $u_{2}=0$, $\Gamma=0.5$ GHz, (a) $u_{1}/\Omega_{sat}=1$, (b) $u_{1}/\Omega_{sat}=3$, (c) $u_{1}/\Omega_{sat}=6$,
(d) $u_{1}/\Omega_{sat}=9$ and $\Delta=-u_{1}$. The efficiency of the trap occupancy control increases with the ratio $\Omega/J$ and tends to zero as $J\rightarrow0$.\label{fig:Fig3}}
\end{figure}

In Fig. \ref{fig:Fig3}, the probability of finding the electron in the state $|1\rangle$ is obtained as a function of the Rabi frequency $\Omega$ and the tunneling rate $J$ for $\Gamma=0.5$ GHz, $\Delta/u_{1}=-1$, and $u_{1}/\Omega_{sat}$ equal to $1$ (a), $3$ (b), $6$ (c), and $9$ (d). If the Coulomb interaction is not high enough, the optical remote control is inefficient and happens only for a reduced set of
values for $\Omega$ and $J$ [Fig. \ref{fig:Fig3} (a)]. As the difference $|u_{1}-u_{2}|$ is increased [Fig. \ref{fig:Fig3} (b) and (c)], the optical control over the charge qubit becomes feasible, with $P_{1}\approx1$ for some values of $\Omega$ and $J$. For high values of $|u_{1}-u_{2}|$ [Fig. \ref{fig:Fig3} (d)], the ability of controlling the charge
qubit saturates and the map of $P_{1}$ as a function of the Rabi frequency and the tunneling rate suffers only small variations.

In the absence of the Coulomb interaction or for $u_{1}\approx u_{2}$, the two qubits would evolve without the influence of each other, as can be noticed from the Hamiltonian of the system [Eq. (\ref{eq:H})]. However, as the difference between these two variables increases, the temporal evolution of the emitting dipole becomes dependent on the charge qubit and vice versa. In this situation, the resonant frequency of the emitter will be shifted by $u_{1}$or $u_{2}$ depending on the electron position state. If the Rabi frequency $\Omega$ is greater than the saturation Rabi frequency $\Omega_{sat}$, the probability of finding the emitter in its excited state is significant ($1/2$ for $\Omega\gg\Omega_{sat}$), therefore it has greater influence on the evolution of the charge qubit. For a small tunneling rate ($J<\Omega,\,|u_{1}-u_{2}|$), the dynamics of the charge qubit is dominated by its interaction with the emitter. Consequently, for $|u_{1}-u_{2}|$ much greater than $\Omega$ and $J$, it can be remotely controlled through the parameters
determining the dynamics of the emitting dipole, such as Rabi frequency, detuning and radiative decay.

\begin{figure}[th]
\begin{centering}
\includegraphics[width=\linewidth]{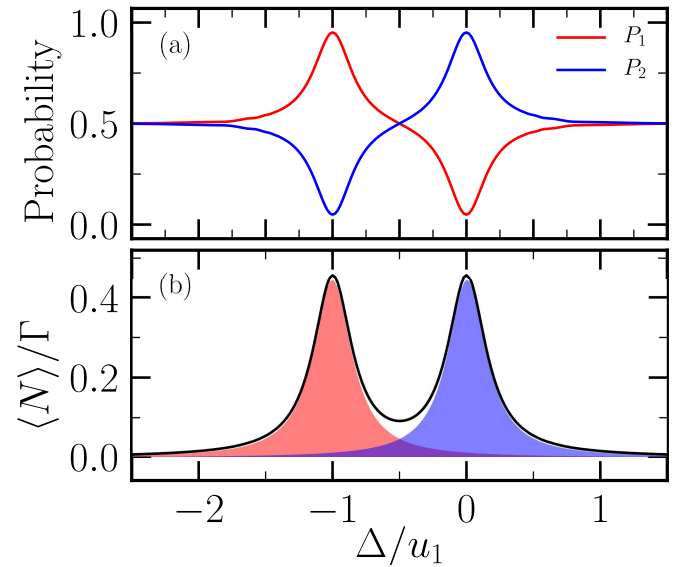}
\par\end{centering}
\caption{(a) Probability of finding the electron in the first trap (red solid line) and second trap (blue solid line) as function of the laser detuning relative to the optical transition of the QD; (b) Expected number of emitted photons $\langle N\rangle$ as a function of the detuning $\Delta$ with the individual contribution of the optical transition
shifted by $u_{1}$ ($u_{2}$) represented by the red (blue) area. The parameters used were $u_{1}/\Omega_{sat}=9$, $u_{2}=0$, $\Omega/\Omega_{sat}=3$, $J/\Omega_{sat}=0.3$ and $\Gamma=0.5$ GHz. \label{fig:Fig4}}
\end{figure}

Taking advantage of the correlation between the two qubits, the photons emitted from the dipole can be used to monitor the electron tunneling between the traps. This is possible because the average photon counting rate \cite{SME_jacobs2006}, given by $\langle N\rangle=\Gamma\langle e|\rho^{s}|e\rangle$, is sensitive to the detuning $\Delta$ and the two possible energy shifts lead to two well separated Lorentzian peaks. The number of photons helps to identify the resonant frequencies, which optimizes the charge qubit control, while the frequency of the photons tells which trap is occupied by the electron (Fig. \ref{fig:Fig4}).

The superposition of the electron position states can also be created by inducing the electron to occupy one of the traps and, in the sequence, eliminating the excitation field. In this case, the population of the electron position states are expected to coherently oscillate with frequency determined by $J$. The evolution of the trap occupancy and, consequently the tunneling rate $J$, can be monitored by applying a relatively small excitation field ($\Omega\ll J$) on resonance with one of the optical transitions. With a spectral distance between the transitions much greater than the linewidth of the Lorentzian peaks, the photon scattering will only happen when the corresponding trap is occupied. A second weak excitation field on resonance with the other energy shift may also be used to complement the monitoring
of the electron.

\begin{figure}[th]
\begin{centering}
\includegraphics[width=\linewidth]{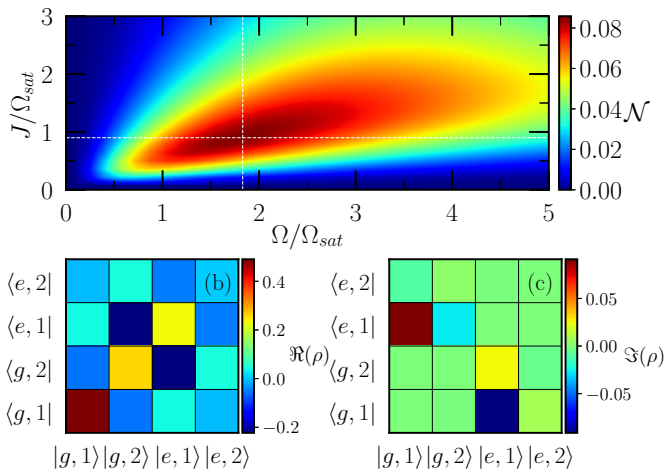}
\par\end{centering}
\caption{(a) Negativity as a function of the ratios $\Omega/\Omega_{sat}$ and $J/\Omega_{sat}$ with $u_{1}=9\Omega_{sat}$, $u_{2}=0$ GHz, $\Gamma=0.5$ GHz and $\Delta=-u_{1}$. Real (b) and imaginary (c) parts of the steady-state density matrix $\rho^{s}$ for $\Omega/\Omega_{sat}=1.8$ and $J/\Omega_{sat}=0.9$, corresponding to the maximal value of the
negativity with $\mathcal{N}\approx0.09$ {[}dashed white line in (a){]}.\label{fig:Fig5}}
\end{figure}

In order to quantify the entanglement between the emitting dipole and the electron position, we analyze the negativity $\mathcal{N}$ defined as
\begin{equation}
\mathcal{N}=\sum_{\lambda<0}|\lambda|,\label{eq:negativity}
\end{equation}
where $\lambda$ are the eigenvalues of the partially transposed density matrix \cite{Negativity_1,negativity_2}. The negativity varies from $0$ for separable states until $1/2$ for fully entangled states. For high values of $\Omega/J$, the states of the emitter have similar populations and modest coherence elements, while the electron tends to occupy one of the traps. In this case, the output from the electron
position measurement has no information on the emitter's state and the two two-level systems are not entangled {[}Fig. \ref{fig:Fig5} (a){]}. A moderate entanglement is obtained by trading-off between the certainty of the electron position and the photon coherence through the decrease of the excitation power. The negativity indicates that a Rabi frequency between $\Omega_{sat}$ and $3\Omega_{sat}$ and
a tunneling rate from $\Omega_{sat}/2$ until $3\Omega_{sat}/2$ favor the entanglement between the emitting dipole and the electron position for $|u_{1}-u_{2}|$ much greater than $\Omega_{sat}$. In this situation, the dynamics of the bipartite system is dominated by the photon emission with the electron occupying the first trap (for $\Delta=-u_{1}$), and by the entangled state of the type $|\psi\rangle=a|g,2\rangle+b|e,1\rangle$, where $a$ and $b$ are complex constants. The fully entangled fraction, defined as
\begin{equation}
\mathcal{F}(\rho^{s})=\max_{\psi}\langle\psi|\rho^{s}|\psi\rangle,\label{eq:entanglement_fraction}
\end{equation}
where $|\psi\rangle$ are all the maximally entangled states of the system, gives a measure of how the mixed steady state $\rho^{s}$ approaches a Bell state \cite{ent_fraction_grondalski2002,ent_fraction_albeverio2002}. For $\Omega/\Omega_{sat}=1.8$, $J/\Omega_{sat}=0.9$ and $u_{1}/\Omega_{sat}=9$ [white lines in Fig. \ref{fig:Fig5} (a)], we have $\mathcal{F}=0.47$ with $|\psi\rangle=\left(|g,2\rangle-|e,1\rangle\right)/\sqrt{2}$ and the negativity is $\mathcal{N}\approx0.09$, which is the maximal value of this map and it is greater than the values expected for the thermal states of a gas-type system \cite{neg_hartmann2007}, for example. In Figs. \ref{fig:Fig5} (b) and \ref{fig:Fig5}(c), we can observe the real and the imaginary parts of the steady-state density matrix $\rho^{s}$, respectively, for the parameters already specified.

The radiative decay of the emitter degrades the quantum correlation between the qubits, however, the entangled state can be distilled if some copies of the system are available \cite{distillation_1,distillation_2,distillation_3,distillation_4,distillation_5,distillation_6,distillation_7}.
Yet, although they are not maximally entangled, the probability of finding the dipole in its excited state and the electron in the second trap is very small ($0.9\%$), while the probability of finding the electron in the first trap (regardless the emitter's state) is about $73\%$, as observed in Figs. \ref{fig:Fig5} (b) and \ref{fig:Fig5} (c).

\section{Conclusion}

In this work, we demonstrated how to remotely control and monitor a single-charge qubit using optical excitation via the Coulomb interaction with an excitonic state of an optically active QD. It was shown that the states of the charge qubit can be manipulated by varying the intensity
and the frequency of the excitation field. Moreover, the control of the electron position state was analyzed for several combinations of the system variables, from where it was concluded that the efficiency tends to unit when $\Omega\gg J$ and $u_{1}\gg\Omega_{sat}$. In this case, the electron position state can be identified through measurements on the amount of photons and their energies. Although the second electron trap was located to give $u_{2}\approx0$, the results presented here are also valid when $u_{2}$ has nonzero values, since it differs from $u_{1}$ enough to resolve the shifted optical transitions, as in Fig. \ref{fig:Fig4} (b).

The entanglement between the optically active qubit and the charge qubit was also investigated and it is present in spite of the radiative decay. When the efficiency of the charge qubit preparation tends to unit, the position of the electron is independent of the optically active qubit and no entanglement is observed. When the Rabi frequency is diminished, the number of photons decreases, the certainty about the electron position becomes smaller, but, in contrast, the entanglement between the qubits competes with the photon emission dynamics. Because the scattered photons carry information about the charge qubit, this system is a candidate for physical implementations in the field of the quantum communication.

A feasible implementation of this system is using solid-state devices, where the phonon-exciton interaction in the optically active qubit may diminish the ability to remotely control the charge qubits, as well as the quantum correlation between them. It can happen because this dephasing mechanism would decrease the spectral resolution of the two transition energies originated from the Coulomb interaction with the electron. However, we do expect these results to still approach reality, given the typical ratio between the quantity of photons emitted at the zero-phonon line and those belonging to the phonon sideband \cite{phonon_forstner2003,phonon_weiler2012}.
\begin{acknowledgments}
Ted S. Santana acknowledges PNPD/CAPES for the support.
\end{acknowledgments}


\begin{thebibliography}{46}%
\makeatletter
\providecommand \@ifxundefined [1]{%
 \@ifx{#1\undefined}
}%
\providecommand \@ifnum [1]{%
 \ifnum #1\expandafter \@firstoftwo
 \else \expandafter \@secondoftwo
 \fi
}%
\providecommand \@ifx [1]{%
 \ifx #1\expandafter \@firstoftwo
 \else \expandafter \@secondoftwo
 \fi
}%
\providecommand \natexlab [1]{#1}%
\providecommand \enquote  [1]{``#1''}%
\providecommand \bibnamefont  [1]{#1}%
\providecommand \bibfnamefont [1]{#1}%
\providecommand \citenamefont [1]{#1}%
\providecommand \href@noop [0]{\@secondoftwo}%
\providecommand \href [0]{\begingroup \@sanitize@url \@href}%
\providecommand \@href[1]{\@@startlink{#1}\@@href}%
\providecommand \@@href[1]{\endgroup#1\@@endlink}%
\providecommand \@sanitize@url [0]{\catcode `\\12\catcode `\$12\catcode
  `\&12\catcode `\#12\catcode `\^12\catcode `\_12\catcode `\%12\relax}%
\providecommand \@@startlink[1]{}%
\providecommand \@@endlink[0]{}%
\providecommand \url  [0]{\begingroup\@sanitize@url \@url }%
\providecommand \@url [1]{\endgroup\@href {#1}{\urlprefix }}%
\providecommand \urlprefix  [0]{URL }%
\providecommand \Eprint [0]{\href }%
\providecommand \doibase [0]{http://dx.doi.org/}%
\providecommand \selectlanguage [0]{\@gobble}%
\providecommand \bibinfo  [0]{\@secondoftwo}%
\providecommand \bibfield  [0]{\@secondoftwo}%
\providecommand \translation [1]{[#1]}%
\providecommand \BibitemOpen [0]{}%
\providecommand \bibitemStop [0]{}%
\providecommand \bibitemNoStop [0]{.\EOS\space}%
\providecommand \EOS [0]{\spacefactor3000\relax}%
\providecommand \BibitemShut  [1]{\csname bibitem#1\endcsname}%
\let\auto@bib@innerbib\@empty
\bibitem [{\citenamefont {Einstein}\ \emph {et~al.}(1935)\citenamefont
  {Einstein}, \citenamefont {Podolsky},\ and\ \citenamefont
  {Rosen}}]{EPS_paradox_1935}%
  \BibitemOpen
  \bibfield  {author} {\bibinfo {author} {\bibfnamefont {A.}~\bibnamefont
  {Einstein}}, \bibinfo {author} {\bibfnamefont {B.}~\bibnamefont {Podolsky}},
  \ and\ \bibinfo {author} {\bibfnamefont {N.}~\bibnamefont {Rosen}},\
  }\href@noop {} {\bibfield  {journal} {\bibinfo  {journal} {Phys. Rev.}\
  }\textbf {\bibinfo {volume} {47}},\ \bibinfo {pages} {777} (\bibinfo {year}
  {1935})}\BibitemShut {NoStop}%
\bibitem [{\citenamefont {Schr\"{o}dinger}(1935)}]{schrodinger_1935}%
  \BibitemOpen
  \bibfield  {author} {\bibinfo {author} {\bibfnamefont {E.}~\bibnamefont
  {Schr\"{o}dinger}},\ }\href {\doibase 10.1017/S0305004100013554} {\bibfield
  {journal} {\bibinfo  {journal} {Math. Proc. Cambridge Philos. Soc.}\ }\textbf
  {\bibinfo {volume} {31}},\ \bibinfo {pages} {555} (\bibinfo {year}
  {1935})}\BibitemShut {NoStop}%
\bibitem [{\citenamefont {Bell}\ and\ \citenamefont
  {Bell}(2004)}]{bell2004speakable}%
  \BibitemOpen
  \bibfield  {author} {\bibinfo {author} {\bibfnamefont {J.~S.}\ \bibnamefont
  {Bell}}\ and\ \bibinfo {author} {\bibfnamefont {J.~S.}\ \bibnamefont
  {Bell}},\ }\href@noop {} {\emph {\bibinfo {title} {Speakable and unspeakable
  in quantum mechanics: Collected papers on quantum philosophy}}}\ (\bibinfo
  {publisher} {Cambridge university press},\ \bibinfo {year}
  {2004})\BibitemShut {NoStop}%
\bibitem [{\citenamefont {Vedral}(2006)}]{vedral2006_book}%
  \BibitemOpen
  \bibfield  {author} {\bibinfo {author} {\bibfnamefont {V.}~\bibnamefont
  {Vedral}},\ }\href@noop {} {\emph {\bibinfo {title} {Introduction to quantum
  information science}}}\ (\bibinfo  {publisher} {Oxford University Press on
  Demand},\ \bibinfo {year} {2006})\BibitemShut {NoStop}%
\bibitem [{\citenamefont {Nielsen}\ and\ \citenamefont
  {Chuang}(2000)}]{nielsen2002quantum}%
  \BibitemOpen
  \bibfield  {author} {\bibinfo {author} {\bibfnamefont {M.~A.}\ \bibnamefont
  {Nielsen}}\ and\ \bibinfo {author} {\bibfnamefont {I.}~\bibnamefont
  {Chuang}},\ }\href@noop {} {\emph {\bibinfo {title} {Quantum computation and
  quantum information}}}\ (\bibinfo  {publisher} {Cambridge University Press,
  Cambridge},\ \bibinfo {year} {2000})\BibitemShut {NoStop}%
\bibitem [{\citenamefont {Kwiat}\ \emph {et~al.}(2001)\citenamefont {Kwiat},
  \citenamefont {Barraza-Lopez}, \citenamefont {Stefanov},\ and\ \citenamefont
  {Gisin}}]{distillation_1}%
  \BibitemOpen
  \bibfield  {author} {\bibinfo {author} {\bibfnamefont {P.~G.}\ \bibnamefont
  {Kwiat}}, \bibinfo {author} {\bibfnamefont {S.}~\bibnamefont
  {Barraza-Lopez}}, \bibinfo {author} {\bibfnamefont {A.}~\bibnamefont
  {Stefanov}}, \ and\ \bibinfo {author} {\bibfnamefont {N.}~\bibnamefont
  {Gisin}},\ }\href@noop {} {\bibfield  {journal} {\bibinfo  {journal}
  {Nature}\ }\textbf {\bibinfo {volume} {409}},\ \bibinfo {pages} {1014}
  (\bibinfo {year} {2001})}\BibitemShut {NoStop}%
\bibitem [{\citenamefont {Takahashi}\ \emph {et~al.}(2010)\citenamefont
  {Takahashi}, \citenamefont {Neergaard-Nielsen}, \citenamefont {Takeuchi},
  \citenamefont {Takeoka}, \citenamefont {Hayasaka}, \citenamefont {Furusawa},\
  and\ \citenamefont {Sasaki}}]{distillation_2}%
  \BibitemOpen
  \bibfield  {author} {\bibinfo {author} {\bibfnamefont {H.}~\bibnamefont
  {Takahashi}}, \bibinfo {author} {\bibfnamefont {J.~S.}\ \bibnamefont
  {Neergaard-Nielsen}}, \bibinfo {author} {\bibfnamefont {M.}~\bibnamefont
  {Takeuchi}}, \bibinfo {author} {\bibfnamefont {M.}~\bibnamefont {Takeoka}},
  \bibinfo {author} {\bibfnamefont {K.}~\bibnamefont {Hayasaka}}, \bibinfo
  {author} {\bibfnamefont {A.}~\bibnamefont {Furusawa}}, \ and\ \bibinfo
  {author} {\bibfnamefont {M.}~\bibnamefont {Sasaki}},\ }\href@noop {}
  {\bibfield  {journal} {\bibinfo  {journal} {Nat. Photonics}\ }\textbf
  {\bibinfo {volume} {4}},\ \bibinfo {pages} {178} (\bibinfo {year}
  {2010})}\BibitemShut {NoStop}%
\bibitem [{\citenamefont {Dong}\ \emph {et~al.}(2008)\citenamefont {Dong},
  \citenamefont {Lassen}, \citenamefont {Heersink}, \citenamefont {Marquardt},
  \citenamefont {Filip}, \citenamefont {Leuchs},\ and\ \citenamefont
  {Andersen}}]{distillation_3}%
  \BibitemOpen
  \bibfield  {author} {\bibinfo {author} {\bibfnamefont {R.}~\bibnamefont
  {Dong}}, \bibinfo {author} {\bibfnamefont {M.}~\bibnamefont {Lassen}},
  \bibinfo {author} {\bibfnamefont {J.}~\bibnamefont {Heersink}}, \bibinfo
  {author} {\bibfnamefont {C.}~\bibnamefont {Marquardt}}, \bibinfo {author}
  {\bibfnamefont {R.}~\bibnamefont {Filip}}, \bibinfo {author} {\bibfnamefont
  {G.}~\bibnamefont {Leuchs}}, \ and\ \bibinfo {author} {\bibfnamefont {U.~L.}\
  \bibnamefont {Andersen}},\ }\href@noop {} {\bibfield  {journal} {\bibinfo
  {journal} {Nat. Phys.}\ }\textbf {\bibinfo {volume} {4}},\ \bibinfo {pages}
  {919} (\bibinfo {year} {2008})}\BibitemShut {NoStop}%
\bibitem [{\citenamefont {Sheng}\ and\ \citenamefont
  {Zhou}(2015)}]{distillation_4}%
  \BibitemOpen
  \bibfield  {author} {\bibinfo {author} {\bibfnamefont {Y.-B.}\ \bibnamefont
  {Sheng}}\ and\ \bibinfo {author} {\bibfnamefont {L.}~\bibnamefont {Zhou}},\
  }\href@noop {} {\bibfield  {journal} {\bibinfo  {journal} {Sci. Rep.}\
  }\textbf {\bibinfo {volume} {5}},\ \bibinfo {pages} {7815} (\bibinfo {year}
  {2015})}\BibitemShut {NoStop}%
\bibitem [{\citenamefont {Hage}\ \emph {et~al.}(2008)\citenamefont {Hage},
  \citenamefont {Samblowski}, \citenamefont {DiGuglielmo}, \citenamefont
  {Franzen}, \citenamefont {Fiur{\'a}{\v{s}}ek},\ and\ \citenamefont
  {Schnabel}}]{distillation_5}%
  \BibitemOpen
  \bibfield  {author} {\bibinfo {author} {\bibfnamefont {B.}~\bibnamefont
  {Hage}}, \bibinfo {author} {\bibfnamefont {A.}~\bibnamefont {Samblowski}},
  \bibinfo {author} {\bibfnamefont {J.}~\bibnamefont {DiGuglielmo}}, \bibinfo
  {author} {\bibfnamefont {A.}~\bibnamefont {Franzen}}, \bibinfo {author}
  {\bibfnamefont {J.}~\bibnamefont {Fiur{\'a}{\v{s}}ek}}, \ and\ \bibinfo
  {author} {\bibfnamefont {R.}~\bibnamefont {Schnabel}},\ }\href@noop {}
  {\bibfield  {journal} {\bibinfo  {journal} {Nat. Phys.}\ }\textbf {\bibinfo
  {volume} {4}},\ \bibinfo {pages} {915} (\bibinfo {year} {2008})}\BibitemShut
  {NoStop}%
\bibitem [{\citenamefont {Vollbrecht}\ \emph {et~al.}(2011)\citenamefont
  {Vollbrecht}, \citenamefont {Muschik},\ and\ \citenamefont
  {Cirac}}]{distillation_6}%
  \BibitemOpen
  \bibfield  {author} {\bibinfo {author} {\bibfnamefont {K.~G.~H.}\
  \bibnamefont {Vollbrecht}}, \bibinfo {author} {\bibfnamefont {C.~A.}\
  \bibnamefont {Muschik}}, \ and\ \bibinfo {author} {\bibfnamefont {J.~I.}\
  \bibnamefont {Cirac}},\ }\href@noop {} {\bibfield  {journal} {\bibinfo
  {journal} {Phys. Rev. Lett.}\ }\textbf {\bibinfo {volume} {107}},\ \bibinfo
  {pages} {120502} (\bibinfo {year} {2011})}\BibitemShut {NoStop}%
\bibitem [{\citenamefont {Datta}\ \emph {et~al.}(2012)\citenamefont {Datta},
  \citenamefont {Zhang}, \citenamefont {Nunn}, \citenamefont {Langford},
  \citenamefont {Feito}, \citenamefont {Plenio},\ and\ \citenamefont
  {Walmsley}}]{distillation_7}%
  \BibitemOpen
  \bibfield  {author} {\bibinfo {author} {\bibfnamefont {A.}~\bibnamefont
  {Datta}}, \bibinfo {author} {\bibfnamefont {L.}~\bibnamefont {Zhang}},
  \bibinfo {author} {\bibfnamefont {J.}~\bibnamefont {Nunn}}, \bibinfo {author}
  {\bibfnamefont {N.~K.}\ \bibnamefont {Langford}}, \bibinfo {author}
  {\bibfnamefont {A.}~\bibnamefont {Feito}}, \bibinfo {author} {\bibfnamefont
  {M.~B.}\ \bibnamefont {Plenio}}, \ and\ \bibinfo {author} {\bibfnamefont
  {I.~A.}\ \bibnamefont {Walmsley}},\ }\href@noop {} {\bibfield  {journal}
  {\bibinfo  {journal} {Phys. Rev. Lett.}\ }\textbf {\bibinfo {volume} {108}},\
  \bibinfo {pages} {060502} (\bibinfo {year} {2012})}\BibitemShut {NoStop}%
\bibitem [{\citenamefont {Loss}\ and\ \citenamefont
  {DiVincenzo}(1998)}]{loss1998quantum}%
  \BibitemOpen
  \bibfield  {author} {\bibinfo {author} {\bibfnamefont {D.}~\bibnamefont
  {Loss}}\ and\ \bibinfo {author} {\bibfnamefont {D.~P.}\ \bibnamefont
  {DiVincenzo}},\ }\href@noop {} {\bibfield  {journal} {\bibinfo  {journal}
  {Phys. Rev. A}\ }\textbf {\bibinfo {volume} {57}},\ \bibinfo {pages} {120}
  (\bibinfo {year} {1998})}\BibitemShut {NoStop}%
\bibitem [{\citenamefont {Kroutvar}\ \emph {et~al.}(2004)\citenamefont
  {Kroutvar}, \citenamefont {Ducommun}, \citenamefont {Heiss}, \citenamefont
  {Bichler}, \citenamefont {Schuh}, \citenamefont {Abstreiter},\ and\
  \citenamefont {Finley}}]{memory_kroutvar2004}%
  \BibitemOpen
  \bibfield  {author} {\bibinfo {author} {\bibfnamefont {M.}~\bibnamefont
  {Kroutvar}}, \bibinfo {author} {\bibfnamefont {Y.}~\bibnamefont {Ducommun}},
  \bibinfo {author} {\bibfnamefont {D.}~\bibnamefont {Heiss}}, \bibinfo
  {author} {\bibfnamefont {M.}~\bibnamefont {Bichler}}, \bibinfo {author}
  {\bibfnamefont {D.}~\bibnamefont {Schuh}}, \bibinfo {author} {\bibfnamefont
  {G.}~\bibnamefont {Abstreiter}}, \ and\ \bibinfo {author} {\bibfnamefont
  {J.~J.}\ \bibnamefont {Finley}},\ }\href@noop {} {\bibfield  {journal}
  {\bibinfo  {journal} {Nature}\ }\textbf {\bibinfo {volume} {432}},\ \bibinfo
  {pages} {81} (\bibinfo {year} {2004})}\BibitemShut {NoStop}%
\bibitem [{\citenamefont {Wang}\ \emph {et~al.}(2012)\citenamefont {Wang},
  \citenamefont {Song},\ and\ \citenamefont {Long}}]{repeater_wang2012}%
  \BibitemOpen
  \bibfield  {author} {\bibinfo {author} {\bibfnamefont {T.-J.}\ \bibnamefont
  {Wang}}, \bibinfo {author} {\bibfnamefont {S.-Y.}\ \bibnamefont {Song}}, \
  and\ \bibinfo {author} {\bibfnamefont {G.~L.}\ \bibnamefont {Long}},\
  }\href@noop {} {\bibfield  {journal} {\bibinfo  {journal} {Phys. Rev. A}\
  }\textbf {\bibinfo {volume} {85}},\ \bibinfo {pages} {062311} (\bibinfo
  {year} {2012})}\BibitemShut {NoStop}%
\bibitem [{\citenamefont {Vamivakas}\ \emph {et~al.}(2009)\citenamefont
  {Vamivakas}, \citenamefont {Zhao}, \citenamefont {Lu},\ and\ \citenamefont
  {Atat{\"u}re}}]{RF_vamivakas2009}%
  \BibitemOpen
  \bibfield  {author} {\bibinfo {author} {\bibfnamefont {A.~N.}\ \bibnamefont
  {Vamivakas}}, \bibinfo {author} {\bibfnamefont {Y.}~\bibnamefont {Zhao}},
  \bibinfo {author} {\bibfnamefont {C.-Y.}\ \bibnamefont {Lu}}, \ and\ \bibinfo
  {author} {\bibfnamefont {M.}~\bibnamefont {Atat{\"u}re}},\ }\href@noop {}
  {\bibfield  {journal} {\bibinfo  {journal} {Nat. Phys.}\ }\textbf {\bibinfo
  {volume} {5}},\ \bibinfo {pages} {198} (\bibinfo {year} {2009})}\BibitemShut
  {NoStop}%
\bibitem [{\citenamefont {Hayashi}\ \emph {et~al.}(2003)\citenamefont
  {Hayashi}, \citenamefont {Fujisawa}, \citenamefont {Cheong}, \citenamefont
  {Jeong},\ and\ \citenamefont {Hirayama}}]{DQD_hayashi2003_1}%
  \BibitemOpen
  \bibfield  {author} {\bibinfo {author} {\bibfnamefont {T.}~\bibnamefont
  {Hayashi}}, \bibinfo {author} {\bibfnamefont {T.}~\bibnamefont {Fujisawa}},
  \bibinfo {author} {\bibfnamefont {H.-D.}\ \bibnamefont {Cheong}}, \bibinfo
  {author} {\bibfnamefont {Y.~H.}\ \bibnamefont {Jeong}}, \ and\ \bibinfo
  {author} {\bibfnamefont {Y.}~\bibnamefont {Hirayama}},\ }\href@noop {}
  {\bibfield  {journal} {\bibinfo  {journal} {Phys. Rev. Lett.}\ }\textbf
  {\bibinfo {volume} {91}},\ \bibinfo {pages} {226804} (\bibinfo {year}
  {2003})}\BibitemShut {NoStop}%
\bibitem [{\citenamefont {Ono}\ \emph {et~al.}(2002)\citenamefont {Ono},
  \citenamefont {Austing}, \citenamefont {Tokura},\ and\ \citenamefont
  {Tarucha}}]{DQD_ono2002_2}%
  \BibitemOpen
  \bibfield  {author} {\bibinfo {author} {\bibfnamefont {K.}~\bibnamefont
  {Ono}}, \bibinfo {author} {\bibfnamefont {D.}~\bibnamefont {Austing}},
  \bibinfo {author} {\bibfnamefont {Y.}~\bibnamefont {Tokura}}, \ and\ \bibinfo
  {author} {\bibfnamefont {S.}~\bibnamefont {Tarucha}},\ }\href@noop {}
  {\bibfield  {journal} {\bibinfo  {journal} {Science}\ }\textbf {\bibinfo
  {volume} {297}},\ \bibinfo {pages} {1313} (\bibinfo {year}
  {2002})}\BibitemShut {NoStop}%
\bibitem [{\citenamefont {Gorman}\ \emph {et~al.}(2005)\citenamefont {Gorman},
  \citenamefont {Hasko},\ and\ \citenamefont {Williams}}]{DQD_gorman2005_4}%
  \BibitemOpen
  \bibfield  {author} {\bibinfo {author} {\bibfnamefont {J.}~\bibnamefont
  {Gorman}}, \bibinfo {author} {\bibfnamefont {D.~G.}~\bibnamefont {Hasko}}, \
  and\ \bibinfo {author} {\bibfnamefont {D.~A.}~\bibnamefont {Williams}},\
  }\href@noop {} {\bibfield  {journal} {\bibinfo  {journal} {Phys. Rev. Lett.}\
  }\textbf {\bibinfo {volume} {95}},\ \bibinfo {pages} {090502} (\bibinfo
  {year} {2005})}\BibitemShut {NoStop}%
\bibitem [{\citenamefont {Englund}\ \emph {et~al.}(2005)\citenamefont
  {Englund}, \citenamefont {Fattal}, \citenamefont {Waks}, \citenamefont
  {Solomon}, \citenamefont {Zhang}, \citenamefont {Nakaoka}, \citenamefont
  {Arakawa}, \citenamefont {Yamamoto},\ and\ \citenamefont
  {Vu{\v{c}}kovi{\'c}}}]{coherence_englund2005}%
  \BibitemOpen
  \bibfield  {author} {\bibinfo {author} {\bibfnamefont {D.}~\bibnamefont
  {Englund}}, \bibinfo {author} {\bibfnamefont {D.}~\bibnamefont {Fattal}},
  \bibinfo {author} {\bibfnamefont {E.}~\bibnamefont {Waks}}, \bibinfo {author}
  {\bibfnamefont {G.}~\bibnamefont {Solomon}}, \bibinfo {author} {\bibfnamefont
  {B.}~\bibnamefont {Zhang}}, \bibinfo {author} {\bibfnamefont
  {T.}~\bibnamefont {Nakaoka}}, \bibinfo {author} {\bibfnamefont
  {Y.}~\bibnamefont {Arakawa}}, \bibinfo {author} {\bibfnamefont
  {Y.}~\bibnamefont {Yamamoto}}, \ and\ \bibinfo {author} {\bibfnamefont
  {J.}~\bibnamefont {Vu{\v{c}}kovi{\'c}}},\ }\href@noop {} {\bibfield
  {journal} {\bibinfo  {journal} {Phys. Rev. Lett.}\ }\textbf {\bibinfo
  {volume} {95}},\ \bibinfo {pages} {013904} (\bibinfo {year}
  {2005})}\BibitemShut {NoStop}%
\bibitem [{\citenamefont {Hennessy}\ \emph {et~al.}(2007)\citenamefont
  {Hennessy}, \citenamefont {Badolato}, \citenamefont {Winger}, \citenamefont
  {Gerace}, \citenamefont {Atat{\"u}re}, \citenamefont {Gulde}, \citenamefont
  {F{\"a}lt}, \citenamefont {Hu},\ and\ \citenamefont
  {Imamo{\u{g}}lu}}]{coherence_hennessy2007}%
  \BibitemOpen
  \bibfield  {author} {\bibinfo {author} {\bibfnamefont {K.}~\bibnamefont
  {Hennessy}}, \bibinfo {author} {\bibfnamefont {A.}~\bibnamefont {Badolato}},
  \bibinfo {author} {\bibfnamefont {M.}~\bibnamefont {Winger}}, \bibinfo
  {author} {\bibfnamefont {D.}~\bibnamefont {Gerace}}, \bibinfo {author}
  {\bibfnamefont {M.}~\bibnamefont {Atat{\"u}re}}, \bibinfo {author}
  {\bibfnamefont {S.}~\bibnamefont {Gulde}}, \bibinfo {author} {\bibfnamefont
  {S.}~\bibnamefont {F{\"a}lt}}, \bibinfo {author} {\bibfnamefont {E.~L.}\
  \bibnamefont {Hu}}, \ and\ \bibinfo {author} {\bibfnamefont {A.}~\bibnamefont
  {Imamo{\u{g}}lu}},\ }\href@noop {} {\bibfield  {journal} {\bibinfo  {journal}
  {Nature}\ }\textbf {\bibinfo {volume} {445}},\ \bibinfo {pages} {896}
  (\bibinfo {year} {2007})}\BibitemShut {NoStop}%
\bibitem [{\citenamefont {Kuhlmann}\ \emph {et~al.}(2015)\citenamefont
  {Kuhlmann}, \citenamefont {Prechtel}, \citenamefont {Houel}, \citenamefont
  {Ludwig}, \citenamefont {Reuter}, \citenamefont {Wieck},\ and\ \citenamefont
  {Warburton}}]{coherence_kuhlmann2015}%
  \BibitemOpen
  \bibfield  {author} {\bibinfo {author} {\bibfnamefont {A.~V.}\ \bibnamefont
  {Kuhlmann}}, \bibinfo {author} {\bibfnamefont {J.~H.}\ \bibnamefont
  {Prechtel}}, \bibinfo {author} {\bibfnamefont {J.}~\bibnamefont {Houel}},
  \bibinfo {author} {\bibfnamefont {A.}~\bibnamefont {Ludwig}}, \bibinfo
  {author} {\bibfnamefont {D.}~\bibnamefont {Reuter}}, \bibinfo {author}
  {\bibfnamefont {A.~D.}\ \bibnamefont {Wieck}}, \ and\ \bibinfo {author}
  {\bibfnamefont {R.~J.}\ \bibnamefont {Warburton}},\ }\href@noop {} {\bibfield
   {journal} {\bibinfo  {journal} {Nat. Commun.}\ }\textbf {\bibinfo {volume}
  {6}},\ \bibinfo {pages} {8204} (\bibinfo {year} {2015})}\BibitemShut
  {NoStop}%
\bibitem [{\citenamefont {Nguyen}\ \emph {et~al.}(2011)\citenamefont {Nguyen},
  \citenamefont {Sallen}, \citenamefont {Voisin}, \citenamefont {Roussignol},
  \citenamefont {Diederichs},\ and\ \citenamefont
  {Cassabois}}]{coherence_nguyen2011}%
  \BibitemOpen
  \bibfield  {author} {\bibinfo {author} {\bibfnamefont {H.-S.}\ \bibnamefont
  {Nguyen}}, \bibinfo {author} {\bibfnamefont {G.}~\bibnamefont {Sallen}},
  \bibinfo {author} {\bibfnamefont {C.}~\bibnamefont {Voisin}}, \bibinfo
  {author} {\bibfnamefont {P.}~\bibnamefont {Roussignol}}, \bibinfo {author}
  {\bibfnamefont {C.}~\bibnamefont {Diederichs}}, \ and\ \bibinfo {author}
  {\bibfnamefont {G.}~\bibnamefont {Cassabois}},\ }\href@noop {} {\bibfield
  {journal} {\bibinfo  {journal} {Appl. Phys. Lett.}\ }\textbf {\bibinfo
  {volume} {99}},\ \bibinfo {pages} {261904} (\bibinfo {year}
  {2011})}\BibitemShut {NoStop}%
\bibitem [{\citenamefont {Flatten}\ \emph {et~al.}(2018)\citenamefont
  {Flatten}, \citenamefont {Weng}, \citenamefont {Branny}, \citenamefont
  {Johnson}, \citenamefont {Dolan}, \citenamefont {Trichet}, \citenamefont
  {Gerardot},\ and\ \citenamefont {Smith}}]{coherence_2D_flatten2018}%
  \BibitemOpen
  \bibfield  {author} {\bibinfo {author} {\bibfnamefont {L.}~\bibnamefont
  {Flatten}}, \bibinfo {author} {\bibfnamefont {L.}~\bibnamefont {Weng}},
  \bibinfo {author} {\bibfnamefont {A.}~\bibnamefont {Branny}}, \bibinfo
  {author} {\bibfnamefont {S.}~\bibnamefont {Johnson}}, \bibinfo {author}
  {\bibfnamefont {P.}~\bibnamefont {Dolan}}, \bibinfo {author} {\bibfnamefont
  {A.}~\bibnamefont {Trichet}}, \bibinfo {author} {\bibfnamefont
  {B.}~\bibnamefont {Gerardot}}, \ and\ \bibinfo {author} {\bibfnamefont
  {J.}~\bibnamefont {Smith}},\ }\href@noop {} {\bibfield  {journal} {\bibinfo
  {journal} {Appl. Phys. Lett.}\ }\textbf {\bibinfo {volume} {112}},\ \bibinfo
  {pages} {191105} (\bibinfo {year} {2018})}\BibitemShut {NoStop}%
\bibitem [{\citenamefont {Claudon}\ \emph {et~al.}(2010)\citenamefont
  {Claudon}, \citenamefont {Bleuse}, \citenamefont {Malik}, \citenamefont
  {Bazin}, \citenamefont {Jaffrennou}, \citenamefont {Gregersen}, \citenamefont
  {Sauvan}, \citenamefont {Lalanne},\ and\ \citenamefont
  {G{\'e}rard}}]{HOM_claudon2010}%
  \BibitemOpen
  \bibfield  {author} {\bibinfo {author} {\bibfnamefont {J.}~\bibnamefont
  {Claudon}}, \bibinfo {author} {\bibfnamefont {J.}~\bibnamefont {Bleuse}},
  \bibinfo {author} {\bibfnamefont {N.~S.}\ \bibnamefont {Malik}}, \bibinfo
  {author} {\bibfnamefont {M.}~\bibnamefont {Bazin}}, \bibinfo {author}
  {\bibfnamefont {P.}~\bibnamefont {Jaffrennou}}, \bibinfo {author}
  {\bibfnamefont {N.}~\bibnamefont {Gregersen}}, \bibinfo {author}
  {\bibfnamefont {C.}~\bibnamefont {Sauvan}}, \bibinfo {author} {\bibfnamefont
  {P.}~\bibnamefont {Lalanne}}, \ and\ \bibinfo {author} {\bibfnamefont
  {J.-M.}\ \bibnamefont {G{\'e}rard}},\ }\href@noop {} {\bibfield  {journal}
  {\bibinfo  {journal} {Nat. Photonics}\ }\textbf {\bibinfo {volume} {4}},\
  \bibinfo {pages} {174} (\bibinfo {year} {2010})}\BibitemShut {NoStop}%
\bibitem [{\citenamefont {Gschrey}\ \emph {et~al.}(2015)\citenamefont
  {Gschrey}, \citenamefont {Thoma}, \citenamefont {Schnauber}, \citenamefont
  {Seifried}, \citenamefont {Schmidt}, \citenamefont {Wohlfeil}, \citenamefont
  {Kr{\"u}ger}, \citenamefont {Schulze}, \citenamefont {Heindel}, \citenamefont
  {Burger} \emph {et~al.}}]{HOM_gschrey2015}%
  \BibitemOpen
  \bibfield  {author} {\bibinfo {author} {\bibfnamefont {M.}~\bibnamefont
  {Gschrey}}, \bibinfo {author} {\bibfnamefont {A.}~\bibnamefont {Thoma}},
  \bibinfo {author} {\bibfnamefont {P.}~\bibnamefont {Schnauber}}, \bibinfo
  {author} {\bibfnamefont {M.}~\bibnamefont {Seifried}}, \bibinfo {author}
  {\bibfnamefont {R.}~\bibnamefont {Schmidt}}, \bibinfo {author} {\bibfnamefont
  {B.}~\bibnamefont {Wohlfeil}}, \bibinfo {author} {\bibfnamefont
  {L.}~\bibnamefont {Kr{\"u}ger}}, \bibinfo {author} {\bibfnamefont {J.-H.}\
  \bibnamefont {Schulze}}, \bibinfo {author} {\bibfnamefont {T.}~\bibnamefont
  {Heindel}}, \bibinfo {author} {\bibfnamefont {S.}~\bibnamefont {Burger}},
  \emph {et~al.},\ }\href@noop {} {\bibfield  {journal} {\bibinfo  {journal}
  {Nat. Commun.}\ }\textbf {\bibinfo {volume} {6}},\ \bibinfo {pages} {7662}
  (\bibinfo {year} {2015})}\BibitemShut {NoStop}%
\bibitem [{\citenamefont {Laurent}\ \emph {et~al.}(2005)\citenamefont
  {Laurent}, \citenamefont {Varoutsis}, \citenamefont {Le~Gratiet},
  \citenamefont {Lema{\^\i}tre}, \citenamefont {Sagnes}, \citenamefont
  {Raineri}, \citenamefont {Levenson}, \citenamefont {Robert-Philip},\ and\
  \citenamefont {Abram}}]{HOM_laurent2005}%
  \BibitemOpen
  \bibfield  {author} {\bibinfo {author} {\bibfnamefont {S.}~\bibnamefont
  {Laurent}}, \bibinfo {author} {\bibfnamefont {S.}~\bibnamefont {Varoutsis}},
  \bibinfo {author} {\bibfnamefont {L.}~\bibnamefont {Le~Gratiet}}, \bibinfo
  {author} {\bibfnamefont {A.}~\bibnamefont {Lema{\^\i}tre}}, \bibinfo {author}
  {\bibfnamefont {I.}~\bibnamefont {Sagnes}}, \bibinfo {author} {\bibfnamefont
  {F.}~\bibnamefont {Raineri}}, \bibinfo {author} {\bibfnamefont
  {A.}~\bibnamefont {Levenson}}, \bibinfo {author} {\bibfnamefont
  {I.}~\bibnamefont {Robert-Philip}}, \ and\ \bibinfo {author} {\bibfnamefont
  {I.}~\bibnamefont {Abram}},\ }\href@noop {} {\bibfield  {journal} {\bibinfo
  {journal} {Appl. Phys. Lett.}\ }\textbf {\bibinfo {volume} {87}},\ \bibinfo
  {pages} {163107} (\bibinfo {year} {2005})}\BibitemShut {NoStop}%
\bibitem [{\citenamefont {Liu}\ \emph {et~al.}(2018)\citenamefont {Liu},
  \citenamefont {Brash}, \citenamefont {O?Hara}, \citenamefont {Martins},
  \citenamefont {Phillips}, \citenamefont {Coles}, \citenamefont {Royall},
  \citenamefont {Clarke}, \citenamefont {Bentham}, \citenamefont {Prtljaga}
  \emph {et~al.}}]{HOM_liu2018}%
  \BibitemOpen
  \bibfield  {author} {\bibinfo {author} {\bibfnamefont {F.}~\bibnamefont
  {Liu}}, \bibinfo {author} {\bibfnamefont {A.~J.}\ \bibnamefont {Brash}},
  \bibinfo {author} {\bibfnamefont {J.}~\bibnamefont {O?Hara}}, \bibinfo
  {author} {\bibfnamefont {L.~M.}\ \bibnamefont {Martins}}, \bibinfo {author}
  {\bibfnamefont {C.~L.}\ \bibnamefont {Phillips}}, \bibinfo {author}
  {\bibfnamefont {R.~J.}\ \bibnamefont {Coles}}, \bibinfo {author}
  {\bibfnamefont {B.}~\bibnamefont {Royall}}, \bibinfo {author} {\bibfnamefont
  {E.}~\bibnamefont {Clarke}}, \bibinfo {author} {\bibfnamefont
  {C.}~\bibnamefont {Bentham}}, \bibinfo {author} {\bibfnamefont
  {N.}~\bibnamefont {Prtljaga}},  \emph {et~al.},\ }\href@noop {} {\bibfield
  {journal} {\bibinfo  {journal} {Nat. Nanotechnol.}\ ,\ \bibinfo {pages} {1}}
  (\bibinfo {year} {2018})}\BibitemShut {NoStop}%
\bibitem [{\citenamefont {Patel}\ \emph {et~al.}(2010)\citenamefont {Patel},
  \citenamefont {Bennett}, \citenamefont {Farrer}, \citenamefont {Nicoll},
  \citenamefont {Ritchie},\ and\ \citenamefont {Shields}}]{HOM_patel2010}%
  \BibitemOpen
  \bibfield  {author} {\bibinfo {author} {\bibfnamefont {R.~B.}\ \bibnamefont
  {Patel}}, \bibinfo {author} {\bibfnamefont {A.~J.}\ \bibnamefont {Bennett}},
  \bibinfo {author} {\bibfnamefont {I.}~\bibnamefont {Farrer}}, \bibinfo
  {author} {\bibfnamefont {C.~A.}\ \bibnamefont {Nicoll}}, \bibinfo {author}
  {\bibfnamefont {D.~A.}\ \bibnamefont {Ritchie}}, \ and\ \bibinfo {author}
  {\bibfnamefont {A.~J.}\ \bibnamefont {Shields}},\ }\href@noop {} {\bibfield
  {journal} {\bibinfo  {journal} {Nat. Photonics}\ }\textbf {\bibinfo {volume}
  {4}},\ \bibinfo {pages} {632} (\bibinfo {year} {2010})}\BibitemShut {NoStop}%
\bibitem [{\citenamefont {Santori}\ \emph {et~al.}(2002)\citenamefont
  {Santori}, \citenamefont {Fattal}, \citenamefont {Vu{\v{c}}kovi{\'c}},
  \citenamefont {Solomon},\ and\ \citenamefont {Yamamoto}}]{HOM_santori2002}%
  \BibitemOpen
  \bibfield  {author} {\bibinfo {author} {\bibfnamefont {C.}~\bibnamefont
  {Santori}}, \bibinfo {author} {\bibfnamefont {D.}~\bibnamefont {Fattal}},
  \bibinfo {author} {\bibfnamefont {J.}~\bibnamefont {Vu{\v{c}}kovi{\'c}}},
  \bibinfo {author} {\bibfnamefont {G.~S.}\ \bibnamefont {Solomon}}, \ and\
  \bibinfo {author} {\bibfnamefont {Y.}~\bibnamefont {Yamamoto}},\ }\href@noop
  {} {\bibfield  {journal} {\bibinfo  {journal} {Nature}\ }\textbf {\bibinfo
  {volume} {419}},\ \bibinfo {pages} {594} (\bibinfo {year}
  {2002})}\BibitemShut {NoStop}%
\bibitem [{\citenamefont {Santana}\ \emph {et~al.}(2017)\citenamefont
  {Santana}, \citenamefont {Ma}, \citenamefont {Malein}, \citenamefont
  {Bastiman}, \citenamefont {Clarke},\ and\ \citenamefont
  {Gerardot}}]{HOM_Ted_2017}%
  \BibitemOpen
  \bibfield  {author} {\bibinfo {author} {\bibfnamefont {T.~S.}\ \bibnamefont
  {Santana}}, \bibinfo {author} {\bibfnamefont {Y.}~\bibnamefont {Ma}},
  \bibinfo {author} {\bibfnamefont {R.~N.~E.}\ \bibnamefont {Malein}}, \bibinfo
  {author} {\bibfnamefont {F.}~\bibnamefont {Bastiman}}, \bibinfo {author}
  {\bibfnamefont {E.}~\bibnamefont {Clarke}}, \ and\ \bibinfo {author}
  {\bibfnamefont {B.~D.}\ \bibnamefont {Gerardot}},\ }\href {\doibase
  10.1103/PhysRevB.95.201410} {\bibfield  {journal} {\bibinfo  {journal} {Phys.
  Rev. B}\ }\textbf {\bibinfo {volume} {95}},\ \bibinfo {pages} {201410}
  (\bibinfo {year} {2017})}\BibitemShut {NoStop}%
\bibitem [{\citenamefont {Spring}\ \emph {et~al.}(2012)\citenamefont {Spring},
  \citenamefont {Metcalf}, \citenamefont {Humphreys}, \citenamefont
  {Kolthammer}, \citenamefont {Jin}, \citenamefont {Barbieri}, \citenamefont
  {Datta}, \citenamefont {Thomas-Peter}, \citenamefont {Langford},
  \citenamefont {Kundys} \emph {et~al.}}]{BS_spring2012}%
  \BibitemOpen
  \bibfield  {author} {\bibinfo {author} {\bibfnamefont {J.~B.}\ \bibnamefont
  {Spring}}, \bibinfo {author} {\bibfnamefont {B.~J.}\ \bibnamefont {Metcalf}},
  \bibinfo {author} {\bibfnamefont {P.~C.}\ \bibnamefont {Humphreys}}, \bibinfo
  {author} {\bibfnamefont {W.~S.}\ \bibnamefont {Kolthammer}}, \bibinfo
  {author} {\bibfnamefont {X.-M.}\ \bibnamefont {Jin}}, \bibinfo {author}
  {\bibfnamefont {M.}~\bibnamefont {Barbieri}}, \bibinfo {author}
  {\bibfnamefont {A.}~\bibnamefont {Datta}}, \bibinfo {author} {\bibfnamefont
  {N.}~\bibnamefont {Thomas-Peter}}, \bibinfo {author} {\bibfnamefont {N.~K.}\
  \bibnamefont {Langford}}, \bibinfo {author} {\bibfnamefont {D.}~\bibnamefont
  {Kundys}},  \emph {et~al.},\ }\href@noop {} {\bibfield  {journal} {\bibinfo
  {journal} {Science}\ ,\ \bibinfo {pages} {1231692}} (\bibinfo {year}
  {2012})}\BibitemShut {NoStop}%
\bibitem [{\citenamefont {Wang}\ \emph {et~al.}(2017)\citenamefont {Wang},
  \citenamefont {He}, \citenamefont {Li}, \citenamefont {Su}, \citenamefont
  {Li}, \citenamefont {Huang}, \citenamefont {Ding}, \citenamefont {Chen},
  \citenamefont {Liu}, \citenamefont {Qin} \emph {et~al.}}]{BS_wang2017}%
  \BibitemOpen
  \bibfield  {author} {\bibinfo {author} {\bibfnamefont {H.}~\bibnamefont
  {Wang}}, \bibinfo {author} {\bibfnamefont {Y.}~\bibnamefont {He}}, \bibinfo
  {author} {\bibfnamefont {Y.-H.}\ \bibnamefont {Li}}, \bibinfo {author}
  {\bibfnamefont {Z.-E.}\ \bibnamefont {Su}}, \bibinfo {author} {\bibfnamefont
  {B.}~\bibnamefont {Li}}, \bibinfo {author} {\bibfnamefont {H.-L.}\
  \bibnamefont {Huang}}, \bibinfo {author} {\bibfnamefont {X.}~\bibnamefont
  {Ding}}, \bibinfo {author} {\bibfnamefont {M.-C.}\ \bibnamefont {Chen}},
  \bibinfo {author} {\bibfnamefont {C.}~\bibnamefont {Liu}}, \bibinfo {author}
  {\bibfnamefont {J.}~\bibnamefont {Qin}},  \emph {et~al.},\ }\href@noop {}
  {\bibfield  {journal} {\bibinfo  {journal} {Nat. Photonics}\ }\textbf
  {\bibinfo {volume} {11}},\ \bibinfo {pages} {361} (\bibinfo {year}
  {2017})}\BibitemShut {NoStop}%
\bibitem [{\citenamefont {DiVincenzo}(2000)}]{criteria_divincenzo2000}%
  \BibitemOpen
  \bibfield  {author} {\bibinfo {author} {\bibfnamefont {D.~P.}\ \bibnamefont
  {DiVincenzo}},\ }\href@noop {} {\bibfield  {journal} {\bibinfo  {journal}
  {Fortschritte der Physik: Progress of Physics}\ }\textbf {\bibinfo {volume}
  {48}},\ \bibinfo {pages} {771} (\bibinfo {year} {2000})}\BibitemShut
  {NoStop}%
\bibitem [{\citenamefont {Mollow}(1969)}]{mollow1969power}%
  \BibitemOpen
  \bibfield  {author} {\bibinfo {author} {\bibfnamefont {B.}~\bibnamefont
  {Mollow}},\ }\href@noop {} {\bibfield  {journal} {\bibinfo  {journal} {Phys.
  Rev.}\ }\textbf {\bibinfo {volume} {188}} (\bibinfo {year}
  {1969})}\BibitemShut {NoStop}%
\bibitem [{\citenamefont {Breuer}\ \emph {et~al.}(2002)\citenamefont {Breuer},
  \citenamefont {Petruccione} \emph {et~al.}}]{Lindblad_breuer2002}%
  \BibitemOpen
  \bibfield  {author} {\bibinfo {author} {\bibfnamefont {H.-P.}\ \bibnamefont
  {Breuer}}, \bibinfo {author} {\bibfnamefont {F.}~\bibnamefont {Petruccione}},
   \emph {et~al.},\ }\href@noop {} {\emph {\bibinfo {title} {The theory of open
  quantum systems}}}\ (\bibinfo  {publisher} {Oxford University Press on
  Demand},\ \bibinfo {year} {2002})\BibitemShut {NoStop}%
\bibitem [{\citenamefont {Press}\ \emph {et~al.}(2007)\citenamefont {Press},
  \citenamefont {Teukolsky}, \citenamefont {Vetterling},\ and\ \citenamefont
  {Flannery}}]{numerical_press2007}%
  \BibitemOpen
  \bibfield  {author} {\bibinfo {author} {\bibfnamefont {W.~H.}\ \bibnamefont
  {Press}}, \bibinfo {author} {\bibfnamefont {S.~A.}\ \bibnamefont
  {Teukolsky}}, \bibinfo {author} {\bibfnamefont {W.~T.}\ \bibnamefont
  {Vetterling}}, \ and\ \bibinfo {author} {\bibfnamefont {B.~P.}\ \bibnamefont
  {Flannery}},\ }\href@noop {} {\emph {\bibinfo {title} {Numerical recipes 3rd
  edition: The art of scientific computing}}}\ (\bibinfo  {publisher}
  {Cambridge university press},\ \bibinfo {year} {2007})\BibitemShut {NoStop}%
\bibitem [{\citenamefont {Knight}\ and\ \citenamefont
  {Milonni}(1980)}]{rabi_knight1980}%
  \BibitemOpen
  \bibfield  {author} {\bibinfo {author} {\bibfnamefont {P.~L.}\ \bibnamefont
  {Knight}}\ and\ \bibinfo {author} {\bibfnamefont {P.~W.}\ \bibnamefont
  {Milonni}},\ }\href@noop {} {\bibfield  {journal} {\bibinfo  {journal} {Phys.
  Rep.}\ }\textbf {\bibinfo {volume} {66}},\ \bibinfo {pages} {21} (\bibinfo
  {year} {1980})}\BibitemShut {NoStop}%
\bibitem [{\citenamefont {Jacobs}\ and\ \citenamefont
  {Steck}(2006)}]{SME_jacobs2006}%
  \BibitemOpen
  \bibfield  {author} {\bibinfo {author} {\bibfnamefont {K.}~\bibnamefont
  {Jacobs}}\ and\ \bibinfo {author} {\bibfnamefont {D.~A.}\ \bibnamefont
  {Steck}},\ }\href@noop {} {\bibfield  {journal} {\bibinfo  {journal}
  {Contemp. Phys.}\ }\textbf {\bibinfo {volume} {47}},\ \bibinfo {pages} {279}
  (\bibinfo {year} {2006})}\BibitemShut {NoStop}%
\bibitem [{\citenamefont {\.{Z}yczkowski}\ \emph {et~al.}(1998)\citenamefont
  {\.{Z}yczkowski}, \citenamefont {Horodecki}, \citenamefont {Sanpera},\ and\
  \citenamefont {Lewenstein}}]{Negativity_1}%
  \BibitemOpen
  \bibfield  {author} {\bibinfo {author} {\bibfnamefont {K.}~\bibnamefont
  {\.{Z}yczkowski}}, \bibinfo {author} {\bibfnamefont {P.}~\bibnamefont
  {Horodecki}}, \bibinfo {author} {\bibfnamefont {A.}~\bibnamefont {Sanpera}},
  \ and\ \bibinfo {author} {\bibfnamefont {M.}~\bibnamefont {Lewenstein}},\
  }\href {\doibase 10.1103/PhysRevA.58.883} {\bibfield  {journal} {\bibinfo
  {journal} {Phys. Rev. A}\ }\textbf {\bibinfo {volume} {58}},\ \bibinfo
  {pages} {883} (\bibinfo {year} {1998})}\BibitemShut {NoStop}%
\bibitem [{\citenamefont {Vidal}\ and\ \citenamefont
  {Werner}(2002)}]{negativity_2}%
  \BibitemOpen
  \bibfield  {author} {\bibinfo {author} {\bibfnamefont {G.}~\bibnamefont
  {Vidal}}\ and\ \bibinfo {author} {\bibfnamefont {R.~F.}\ \bibnamefont
  {Werner}},\ }\href {\doibase 10.1103/PhysRevA.65.032314} {\bibfield
  {journal} {\bibinfo  {journal} {Phys. Rev. A}\ }\textbf {\bibinfo {volume}
  {65}},\ \bibinfo {pages} {032314} (\bibinfo {year} {2002})}\BibitemShut
  {NoStop}%
\bibitem [{\citenamefont {Grondalski}\ \emph {et~al.}(2002)\citenamefont
  {Grondalski}, \citenamefont {Etlinger},\ and\ \citenamefont
  {James}}]{ent_fraction_grondalski2002}%
  \BibitemOpen
  \bibfield  {author} {\bibinfo {author} {\bibfnamefont {J.}~\bibnamefont
  {Grondalski}}, \bibinfo {author} {\bibfnamefont {D.}~\bibnamefont
  {Etlinger}}, \ and\ \bibinfo {author} {\bibfnamefont {D.}~\bibnamefont
  {James}},\ }\href@noop {} {\bibfield  {journal} {\bibinfo  {journal} {Phys.
  Lett. A}\ }\textbf {\bibinfo {volume} {300}},\ \bibinfo {pages} {573}
  (\bibinfo {year} {2002})}\BibitemShut {NoStop}%
\bibitem [{\citenamefont {Albeverio}\ \emph {et~al.}(2002)\citenamefont
  {Albeverio}, \citenamefont {Fei},\ and\ \citenamefont
  {Yang}}]{ent_fraction_albeverio2002}%
  \BibitemOpen
  \bibfield  {author} {\bibinfo {author} {\bibfnamefont {S.}~\bibnamefont
  {Albeverio}}, \bibinfo {author} {\bibfnamefont {S.-M.}\ \bibnamefont {Fei}},
  \ and\ \bibinfo {author} {\bibfnamefont {W.-L.}\ \bibnamefont {Yang}},\
  }\href@noop {} {\bibfield  {journal} {\bibinfo  {journal} {Phys. Rev. A}\
  }\textbf {\bibinfo {volume} {66}},\ \bibinfo {pages} {012301} (\bibinfo
  {year} {2002})}\BibitemShut {NoStop}%
\bibitem [{\citenamefont {Hartmann}\ \emph {et~al.}(2007)\citenamefont
  {Hartmann}, \citenamefont {D{\"u}r},\ and\ \citenamefont
  {Briegel}}]{neg_hartmann2007}%
  \BibitemOpen
  \bibfield  {author} {\bibinfo {author} {\bibfnamefont {L.}~\bibnamefont
  {Hartmann}}, \bibinfo {author} {\bibfnamefont {W.}~\bibnamefont {D{\"u}r}}, \
  and\ \bibinfo {author} {\bibfnamefont {H.}~\bibnamefont {Briegel}},\
  }\href@noop {} {\bibfield  {journal} {\bibinfo  {journal} {New Journal of
  Physics}\ }\textbf {\bibinfo {volume} {9}},\ \bibinfo {pages} {230} (\bibinfo
  {year} {2007})}\BibitemShut {NoStop}%
\bibitem [{\citenamefont {F{\"o}rstner}\ \emph {et~al.}(2003)\citenamefont
  {F{\"o}rstner}, \citenamefont {Weber}, \citenamefont {Danckwerts},\ and\
  \citenamefont {Knorr}}]{phonon_forstner2003}%
  \BibitemOpen
  \bibfield  {author} {\bibinfo {author} {\bibfnamefont {J.}~\bibnamefont
  {F{\"o}rstner}}, \bibinfo {author} {\bibfnamefont {C.}~\bibnamefont {Weber}},
  \bibinfo {author} {\bibfnamefont {J.}~\bibnamefont {Danckwerts}}, \ and\
  \bibinfo {author} {\bibfnamefont {A.}~\bibnamefont {Knorr}},\ }\href@noop {}
  {\bibfield  {journal} {\bibinfo  {journal} {Physical review letters}\
  }\textbf {\bibinfo {volume} {91}},\ \bibinfo {pages} {127401} (\bibinfo
  {year} {2003})}\BibitemShut {NoStop}%
\bibitem [{\citenamefont {Weiler}\ \emph {et~al.}(2012)\citenamefont {Weiler},
  \citenamefont {Ulhaq}, \citenamefont {Ulrich}, \citenamefont {Richter},
  \citenamefont {Jetter}, \citenamefont {Michler}, \citenamefont {Roy},\ and\
  \citenamefont {Hughes}}]{phonon_weiler2012}%
  \BibitemOpen
  \bibfield  {author} {\bibinfo {author} {\bibfnamefont {S.}~\bibnamefont
  {Weiler}}, \bibinfo {author} {\bibfnamefont {A.}~\bibnamefont {Ulhaq}},
  \bibinfo {author} {\bibfnamefont {S.~M.}~\bibnamefont {Ulrich}}, \bibinfo
  {author} {\bibfnamefont {D.}~\bibnamefont {Richter}}, \bibinfo {author}
  {\bibfnamefont {M.}~\bibnamefont {Jetter}}, \bibinfo {author} {\bibfnamefont
  {P.}~\bibnamefont {Michler}}, \bibinfo {author} {\bibfnamefont
  {C.}~\bibnamefont {Roy}}, \ and\ \bibinfo {author} {\bibfnamefont
  {S.}~\bibnamefont {Hughes}},\ }\href@noop {} {\bibfield  {journal} {\bibinfo
  {journal} {Physical Review B}\ }\textbf {\bibinfo {volume} {86}},\ \bibinfo
  {pages} {241304} (\bibinfo {year} {2012})}\BibitemShut {NoStop}%
\end{thebibliography}
\end{document}